\documentclass[aps,prb,preprint,superscriptaddress,showpacs]{revtex4-1}
\usepackage{graphicx}
\usepackage{amsmath}

\usepackage{color}

\begin{document}


\title{Energy Exchange between Phononic and Electronic Subsystem Governing The Nonlinear Conduction in DCNQI$_2$Cu}


\author{Florian Huewe}
\email[]{fhuewe@physik.uni-wuerzburg.de}
\affiliation{Julius-Maximilians-University, Am Hubland, 97074 W\"urzburg, Germany}

\author{Alexander Steeger}
\affiliation{Julius-Maximilians-University, Am Hubland, 97074 W\"urzburg, Germany}

\author{Irene Bauer}
\affiliation{Bayreuth University, Universit\"atsstr. 30, 95447 Bayreuth, Germany}

\author{Steffen Doerrich}
\affiliation{Julius-Maximilians-University, Am Hubland, 97074 W\"urzburg, Germany}

\author{Peter Strohriegl}
\affiliation{Bayreuth University, Universit\"atsstr. 30, 95447 Bayreuth, Germany}

\author{Jens Pflaum}
\affiliation{Julius-Maximilians-University, Am Hubland, 97074 W\"urzburg, Germany}
\affiliation{ZAE Bayern, Am Galgenberg 87, 97074 W\"urzburg, Germany}


\date{\today}

\begin{abstract}

We present a dynamical study on the nonlinear conduction behaviour in the commensurate charge-density-wave phase of the quasi-one-dimensional conductor DCNQI$_2$Cu below 75 K. We can accurately simulate magnitude and time-dependence of the measured conductivity in response to large voltage pulses by accounting for the energy exchange between the phononic and electronic subsystems by means of an electrothermal model. Our simulations reveal a distinct non-equilibrium population of optical phonon states with an average energy of $\overline{E_{ph}}=19$ meV being half the activation energy of about $\Delta E_a= 39$ meV observed in DC resistivity measurements. By inelastic scattering, this hot optical phonon bath generates additional charge-carrying excitations thus providing a multiplication effect while energy transferred to the acoustic phonons is dissipated out of the system via heat conduction. Therefore, in high electric fields a preferred interaction of charge-carrying excitations with optical phonons compared to acoustic phonon modes is considered to be responsible for the nonlinear conduction effects observed in DCNQI$_2$Cu.


\end{abstract}


\maketitle

\section{Introduction}
\label{sec:Intro}

Organic semiconductors have been widely investigated in recent decades with respect to their potential application in electronic devices. Often their electron and hole mobilities are limited due to an efficient interaction of charge carriers with phonons the latter being lower in energy compared to conventional inorganic semiconductors because of the weak intermolecular binding forces. More exotic semiconducting ground states, such as Peierls or Mott insulators, occur in low-dimensional organic conductors often leading to nonlinear conduction phenomena\cite{Iwasa1989, Mori2008}. These nonlinearities seem to be distinctive for materials, otherwise distinguished by quite different ground states, such as spin-Peierls systems like K-TCNQ \cite{Kumai1999} or charge-ordered states as in $\alpha$-(BEDT-TTF)$_2$I$_3$ \cite{Tamura2010, Ivek2012}, and therefore aim for a more general description. Mori et al. \cite{Mori2009} were the first to recognize the connection between the steepness of the resistance curves and the electric threshold fields above which nonlinear conduction occurred. They proposed an electrothermal model to describe the current-voltage characteristics explicitly considering electrical (Joule) heating of distinct electronic or phononic degrees of freedom in the material. In this model, the effective specific heat is the crucial parameter hinting at the mechanism that governs the nonlinear transport effects. This quantity is usually interpreted as electronic specific heat caused by the heating of the electronic system at concurrent inefficient energy transfer between electrons and phonons in contrast to conventional organic semiconductors.\\
To study nonlinear conduction and the validity of the electrothermal model suggested by Mori et al. \cite{Mori2009}, we chose Copper radical anion salts of the Dimethyl-Dicyanoquinonediimine (DCNQI) molecule, first synthesized and studied in the group of Huenig \cite{Aumueller1986} as an archetypical low-dimensional molecular metal. Cu and DCNQI form a quasi-one-dimensional radical anion salt with separate stacking structure staying metallic down to lowest temperatures. In (DCNQI-d$_6$)$_2$Cu the methyl groups are deuterated causing a three-fold commensurate lattice distortion along the face-to-face stacked molecular columns and charge ordering of same periodicity at the Cu sites below around 80K. This phase transition from the metallic to the commensurate charge-density-wave (CDW) ground state is of first order resulting in a step-like increase of the conductivity\cite{Kato1993}. Nonlinear conduction in the CDW state of DCNQI$_2$Cu radical anion salts has been observed before \cite{Wakita2010,Vuletic2001}. By means of dielectric response measurements the excitation of soliton-like charged domain wall pairs has been identified as relaxation channel indicating their importance for the transport properties in the semiconducting phase of DCNQI$_2$Cu \cite{pinterić2000low}.\\
Furthermore, the ground state of deuterated DCNQI$_2$Cu is very sensitive to the change in energy when switching from the free electron gas in the metallic phase to localized electrons in the insulating phase \cite{Nishio2000}, i.e. a gain in electronic energy or delocalization of electrons by an applied electric field should be more likely to induce an insulator-metal transition compared to other systems. Finally, the availability of high-quality literature data\cite{Matsui2009, Nishio2000} on the temperature-dependent contributions of the electronic system as well as of the acoustic and optical phonons to the specific heat allows for an identification of the leading microscopic mechanism causing nonlinear conduction.\\
In this paper we will show that nonlinear conduction in DCNQI$_2$Cu can be well described by means of the above-mentioned electrothermal model. For this purpose we carried out detailed transient analyses of the conduction behavior allowing for a comprehensive determination of the effective specific heat. The magnitude of this quantity and its temperature-dependence rendered it possible to identify its major contribution from optical phonons with an energy of about half the activation energy deduced from the temperature-dependent low-field conductivity. Based on these observations we develop an important extension of the established electrothermal model by including the distinct electrical heating of the optical phonon subsystem requiring an efficient interaction of the charge-carrying excitations with optical phonons similar to conventional organic semiconductors\cite{Karl2003}.
%

\section{Theory}
\label{sec:Theo}

In order to account for the universal nonlinear conduction behavior in a variety of organic charge-transfer materials with different ground states, Mori et al. proposed an electrothermal model\cite{Mori2009}:

\begin{equation}
	nC\frac{dT}{dt}=P+\nabla \cdot\left\{\kappa \nabla T\right\}
	\label{eq:electrothermal-model}
\end{equation}

where $nC$ denotes the effective heat capacity per volume, $P$ is the supplied electrical energy, $\kappa$ is the thermal conductivity and $T$ is the temperature of the system. The supplied electrical energy reads $P=\sigma(T)E^2$, where $\sigma(T)$ is the temperature-dependent conductivity of the material and $E$ is the applied electric field. Explicitly neglecting the spatial variation of T, Eq. \ref{eq:electrothermal-model} can be simplified by a linear expansion in temperature:

\begin{equation}
	nC\frac{dT}{dt}=\sigma(T)E^2-\alpha\left\{T-T_0\right\}
	\label{eq:electrothermal-model-simple}
\end{equation}

\begin{figure}[htb]
	\includegraphics[width=0.5\textwidth]{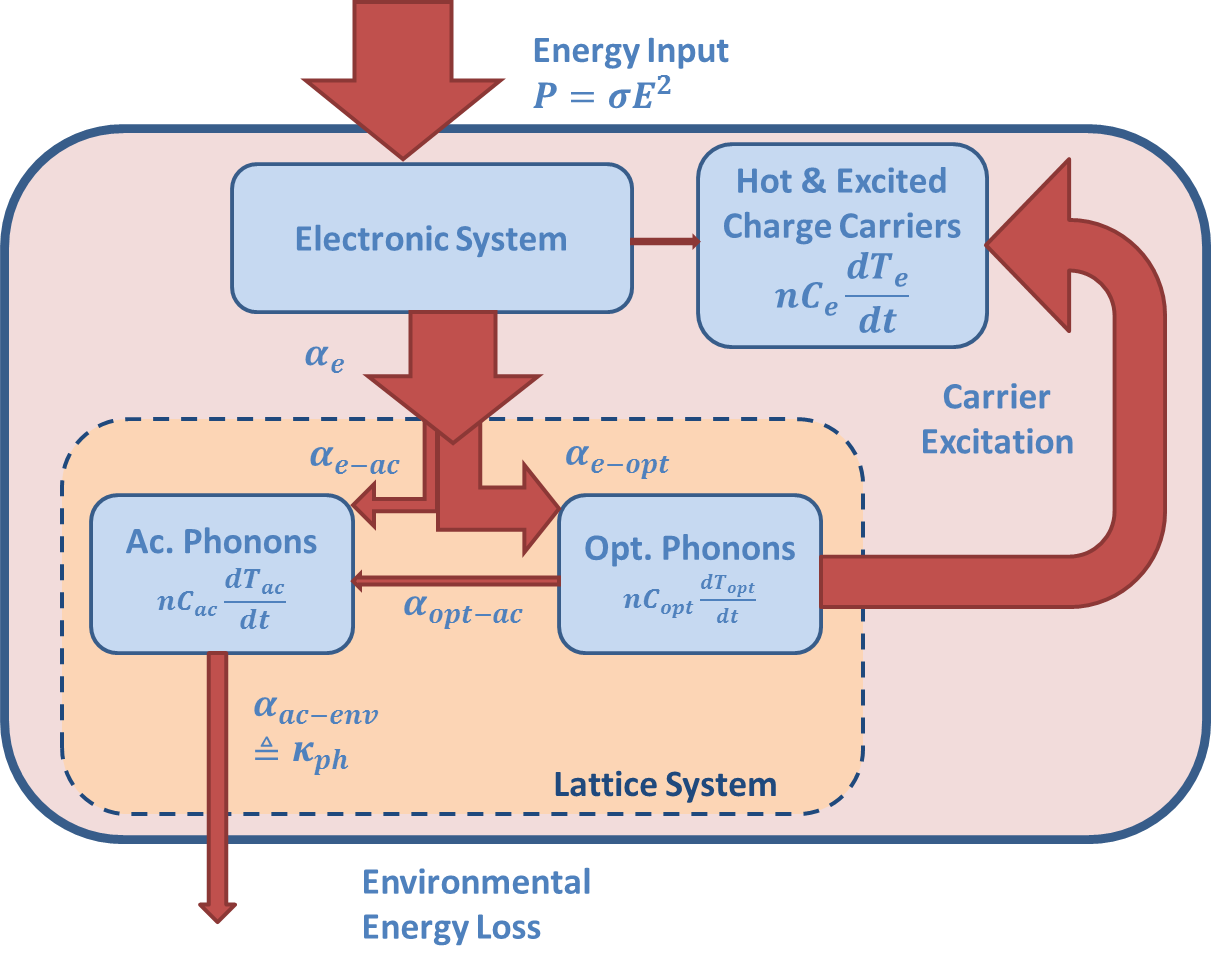}
	\caption{Illustration of the advanced electrothermal model. Depending on the charge carrier scattering rates, the electrical energy can be transferred to distinct microscopic subsystems of the material. An inefficient electron lattice scattering may lead to a direct excitation of charge carriers or hot electrons by the electric field. Due to their low contribution to thermal conductivity, energy stored in the optical phonon system is available for excitation of additional charge carriers  while energy in the acoustic phonon subsystem quickly dissipates out of the sample. Therefore, identification of the microscopic contributions to the effective specific heat, the latter governing the electrothermal model, allows for an understanding of the energy flow in the system in relation to the preferred charge carrier scattering mechanism.}
\label{fig:model}	
\end{figure}

Here, $T_0$ is the ambient temperature and $\alpha$ is an energy transfer rate from the sample to the environment depending on the specific dissipation mechanism and the sample geometry. It has to be emphasized that inhomogeneous temperature and current distributions are neglected throughout this approach. Rather this model describes the electrical heating of a material in presence of just a heat flow out of the sample. By assigning the effective specific heat to distinct microscopic processes of electronic or phononic (optical as well as acoustic) nature, the energy deposition into the individual microscopic subsystems and the dominant scattering mechanism of charge carriers can be analyzed. Therefore, an extended version of the model proposed by Mori is schematically illustrated in Fig. \ref{fig:model} explicitly taking into consideration the separation of the lattice system into the acoustic and optical phonon part and allowing for an additional charge carrier excitation channel via coupling to optical phonons. First of all, the energy provided by the electric field is fed into the electronic system. Depending on the electron scattering rate by acoustic and optical phonons ($\alpha_e=\alpha_{e-ac}+\alpha_{e-opt}$), this energy can be transferred to the lattice system. However, if $\alpha_e$ is small, hot carrier generation or direct population of excited electron states constitute the preferred relaxation mechanisms of the energy supplied by the electric field. This would be indicated by a low electronic contribution $nC_e$ to the specific heat measured in this case. In this case, the temperature $T=T_e$ is assumed to be a parameter describing the population of excited electronic states rather than resembling the crystal lattice temperature $T_L$. On the other hand, identification of distinct contributions to the lattice specific heat $nC_L=nC_{ac}+nC_{opt}$ to $nC$ yields evidence of an efficient carrier scattering by phonons. The relative magnitude of the electron scattering by acoustic and optical phonons also distinguishes the preferred excitation of the respective phononic subsystem. Yet, optical phonons do not contribute significantly to the heat transport because of their low group velocity, i.e. the energy stored in the optical phonon system is available quite long for an excitation of additional charge carriers while the heat stored in acoustic phonons quickly dissipates out of the system. Therefore, the measured $\alpha$ refers to limiting process of the energy transport out of the system.

By numerical integration of Eq. \ref{eq:electrothermal-model-simple}, the transient conductivity $\sigma(T[t])$ of the material for the applied electric field can be determined if the temperature-dependent conductivity is known. The parameters $\alpha$ and $nC$ can be adjusted to match the simulations with the measured nonlinear current-voltage characteristics. To account for the observed nonlinear conduction phenomena in a variety of organic conductors, $nC$ usually is assumed to be small compared to the specific heat of the material and is identified as the electronic specific heat $nC_e$\cite{Mori2008, Mori2009, Wakita2010}. Under these conditions, the temperature-dependence of the threshold electric field $E_{th}$ as well as of the threshold current $j_{th}$ can be remarkably well reproduced. Nevertheless, a detailed analysis of the temperature-dependent effective specific heat as well as of the possible excitation of distinct lattice degrees of freedom accounting for the low $nC$ observed, has not been presented so far. To account for voltage drops at the load resistor in our simulations we adjusted Eq. \ref{eq:electrothermal-model-simple} to the following form 

\begin{equation}
	\frac{dT}{dt}=\underbrace{\frac{m_{mol}\left\{\frac{U}{1+R_L/R_S(T)}\right\}^2}{\rho \cdot V \cdot c_m(T)\cdot R_S(T)}}_{\text{P-term}}-\underbrace{\frac{m_{mol}\cdot \widetilde{\alpha}}{\rho \cdot V \cdot c_m(T)}\left\{T-T_0\right\}}_{\text{K-term}}
	\label{eq:electrothermal-model-adj}
\end{equation}

with the sample volume $V$, the density\cite{Sinzger1993} $\rho=1.61$ g/cm$^3$, the temperature-dependent molar specific heat $c_m(T)$ and the molar mass $m_{mol}$. $U$ is the voltage applied to the series of sample and load resistor ($R_S$ and $R_L$) and $\widetilde{\alpha}$ is the absolute energy transfer rate in $[W/K]$. The first term on the right hand side of Eq. \ref{eq:electrothermal-model-adj} will be denoted by P-term and second one by K-term from now on. Later, we also will allow for the specific heat to become temperature-dependent. \\
By neglecting the K-term and assuming a constant heating power ($R_S=$const), it is possible to extract the specific heat from the linear relation $\Delta T=($P-term$)\cdot \Delta t$ on short time-scales. This method has already been proven to work reasonably well in measuring the specific heat of rod-like samples, e.g. platinum wires\cite{Guo2007}. Yet, the temporal resolution has to be sufficient to neglect the K-term which often is problematic in samples with higher resistance. Therefore, we approximate the K-term for short time-scales and small temperature steps by assuming a constant average temperature difference $\theta=T-T_0=$const being half the temperature increase the transient linear fit extends to. Furthermore, we assume a constant effective heating power by approximating $R_S(T)=R_{S,eff}=R_S(T_0+\theta)$ revealing a simple analytic solution of Eq. \ref{eq:electrothermal-model-adj}

\begin{eqnarray}
	\Delta T =&\left\{\frac{U^2\cdot m_{mol} (1+R_L/R_{S,eff})^{-2}}{\rho V c_m(T_0+\theta)R_{S,eff}} -\frac{m_{mol}\cdot\widetilde{\alpha} \cdot \theta}{\rho V c_m(T_0+\theta)} \right\}\cdot t  \nonumber \\
	=& \left\{B \cdot U^2-C\right\}\cdot t = A_{eff}\cdot t 
	\label{eq:electrothermal-model-lowT}
\end{eqnarray}

From the voltage dependence of the slope $A_{eff} \propto U^2$ the specific heat of the system under study can be determined quite reliably as the K-Term only represents a voltage independent offset. We found that utilizing the specific heat determined by this method allows us to simulate the transient resistance of DCNQI$_2$Cu very accurately.

\section{Experimental}
\label{sec:Exp}

DCNQI-d$_6$ was prepared according to a literature procedure\cite{Hunig1998}. High-quality single crystals of (DCNQI-d$_6$)$_2$Cu were grown electrochemically as described by Kato et al.\cite{Kato1989}. The needle-like crystals with typical cross-section areas of (0.03-0.1)$^2$ mm$^2$ and lengths of (0.2-20) mm were attached to free-standing gold wires of $25$ $\mu$m diameter by silver-paint. Care has been taken to cover the ends of the crystal completely with silver paint to ensure a uniform current injection into the sample.
The temperature-dependent resistance was measured for different crystals from the same batch in two- and four-probe geometry  in a He-flow cryostat between 4 K and 300 K with a Keithley 236 Source Measurement Unit (SMU). The data on nonlinear conduction presented here was measured in two-probe geometry on the same sample with a serial load resistor being roughly two orders of magnitude smaller than the sample resistance at low fields. Pulsed current-voltage characteristics were recorded by applying voltage pulses of 5-40 ms duration separated by 0.5 s intervals and measuring the current. In case of the transient resistance measurements, voltage and current were recorded with a Tektronix TDS744 oscilloscope. The applied voltage was fed into the oscilloscope by a 100:1 passive probe with $R_{in}=$100 M$\Omega$ input impedance while at the load resistor usually a 10:1 passive probe with $R_{in}=$10 M$\Omega$ was used. A sketch of our transient setup is shown in Fig. \ref{fig:Nonlinear-CV-Characteristics}a.

\section{Experimental Results}
\label{sec:Results}

\subsection{Nonlinear Current-Voltage Characteristics}
\label{sec:res-CV}

\begin{figure*}[htb]
	\includegraphics[width=1\textwidth]{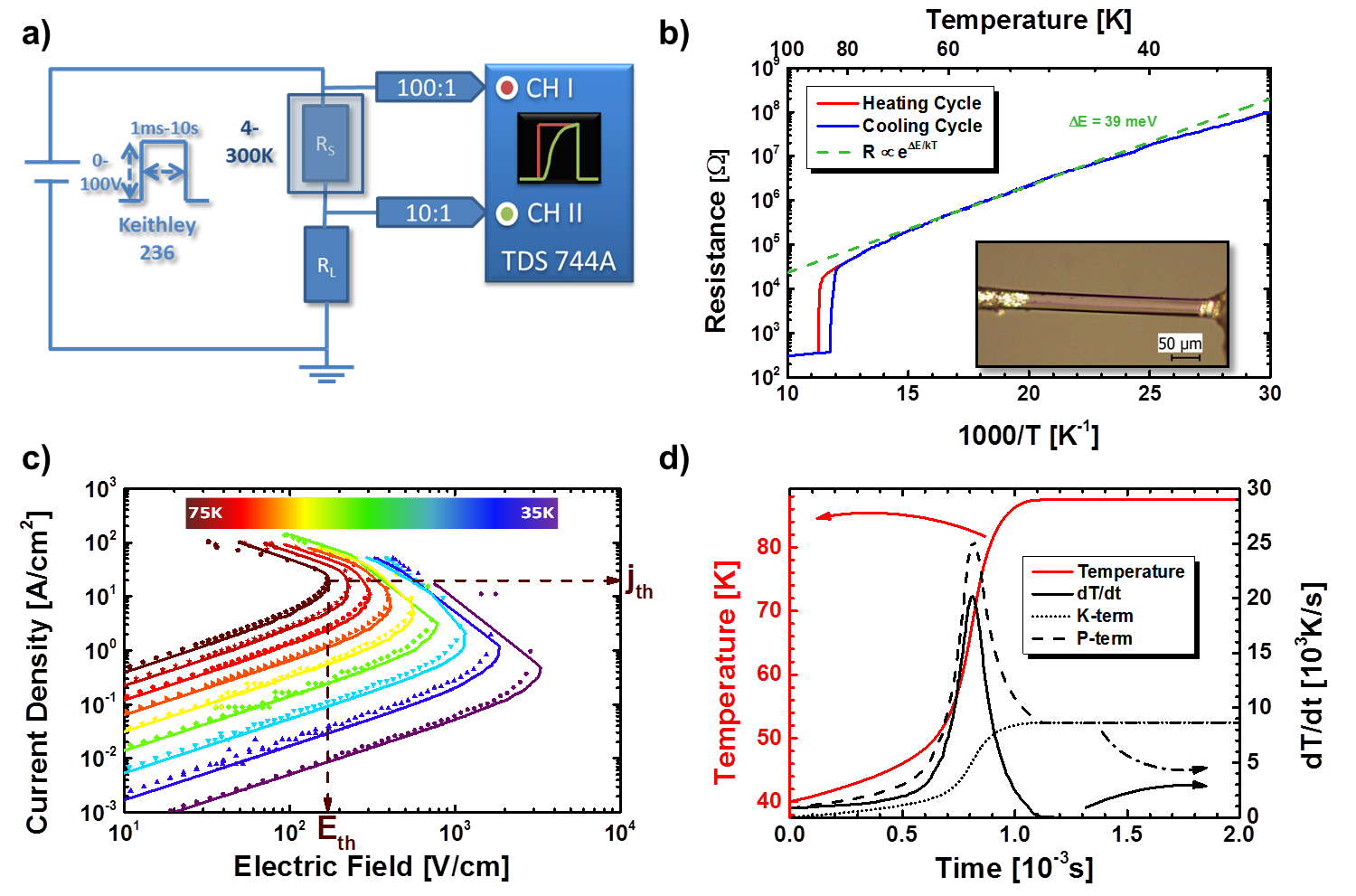}
	\caption{(a) Setup used for the transient resistance experiments. (b) Two-probe sample resistance during the cooling and the heating cycle plotted logarithmically against the inverse temperature. The linear fit yields an activation energy of $\Delta E \approx$ 39 meV below the phase transition. The inset shows a microscope image of the sample. (c) Measured (dots) and simulated (solid lines) nonlinear current-voltage characteristics of (DCNQI-d$_6$)$_2$Cu between 35 K and 75 K in steps of 5 K. (d) Simulated temperature transients after applying a voltage pulse of 87 V to the crystal at $T_0=40$ K. The transient P- and K-terms (Eq. \ref{eq:electrothermal-model-adj}) contributing to the temperature rise dT/dt are illustrated separately.}
\label{fig:Nonlinear-CV-Characteristics}	
\end{figure*}

The obtained crystals always revealed room temperature conductivities of 600-800 S/cm comparable to values in literature\cite{Bauer1993, Kato1993}. The conductivity increased according to the typical $\sigma \propto T^{-2.1}$ dependence down to about $T_{c}=$ 85 K where its drop by five orders of magnitude indicated the first-order Charge-Density-Wave (CDW) transition. Upon heating the transition occurred with a small hysteresis at $T_{c}=$ 89 K typical for first-order phase transitions. As depicted in Fig. \ref{fig:Nonlinear-CV-Characteristics}b, the two-probe resistance measurement only reveals a jump by two orders of magnitude in resistance, i.e. the contact resistance can be neglected below the phase transition but dominates the conduction behavior above $T_c$. Above the phase transition, the the contact resistance shows an activation energy of about 10 meV (fit not shown). Below $T_c$ the resistivity is thermally activated with a corresponding energy of about $\Delta E_a=$ (39$\pm$5) meV in agreement with values obtained from four-probe measurements on various crystals of the same batch. The interpolated two-probe resistance from the heating cycle of the sample with cross-section area of (0.03)$^2$ mm$^2$ and length of (0.275) mm was used for all subsequent simulations.\\
Fig. \ref{fig:Nonlinear-CV-Characteristics}c illustrates the nonlinear current-voltage characteristics obtained from measurements with single voltage pulses of 40 ms length and 0.2-100 V magnitude at temperatures between 35 K and 75 K together with the numerical simulation according to Eq. \ref{eq:electrothermal-model-adj} with $c_m=$ 55 J/(mole K) and $\widetilde{\alpha}=$ 9$\cdot 10^{-5}$ W/K kept constant. The simulation is able to accurately reproduce the current-voltage characteristics and the temperature-dependence of threshold field $E_{th}$ and current $j_{th}$. The reason for the reliability of the simulation can be understood by Fig. \ref{fig:Nonlinear-CV-Characteristics}d where the transient evolution of the temperature obtained from the simulation for a voltage pulse of 87 V at $T_0=$ 40 K is shown. Here, already after 1.1 ms the steady state ($dT/dt=0$) is reached which will be the case for most of the data points in the current-voltage characteristics. As the steady state solution of Eq. \ref{eq:electrothermal-model-adj} is independent of $c_m$, the final temperature and thus the resistance is determined by the choice of $\widetilde{\alpha}$ as long as the value for $c_m$ is chosen not too high. Fig. \ref{fig:Nonlinear-CV-Characteristics}d also displays the individual contributions of the P- and the K-terms to the transient temperature rise $dT/dt$. The P-term governs the short-time behavior until the K-term reacts on the temperature rise leading to its steady state solution. It has to be noted that both terms still depend on $c_m$ in our description, i.e. until the steady state is reached, the dynamic solution depends on $c_m$ and $\widetilde{\alpha}$. The applied value of $c_m=$ 55 J/(mole K) corresponds to the specific heat of (DCNQI-d$_6$)$_2$Cu at 30 K \cite{Matsui2009}. In order to gain detailed insights into the dependence of the resistive switching on microscopic contributions to the specific heat, shorter pulses are required or, even better, the complete transient evolution of the sample resistance has to be analyzed which we will address in the following section. 

\subsection{Dynamic Response}
\label{sec:Res-DR}

\begin{figure*}[htb]
	\includegraphics[width=1\textwidth]{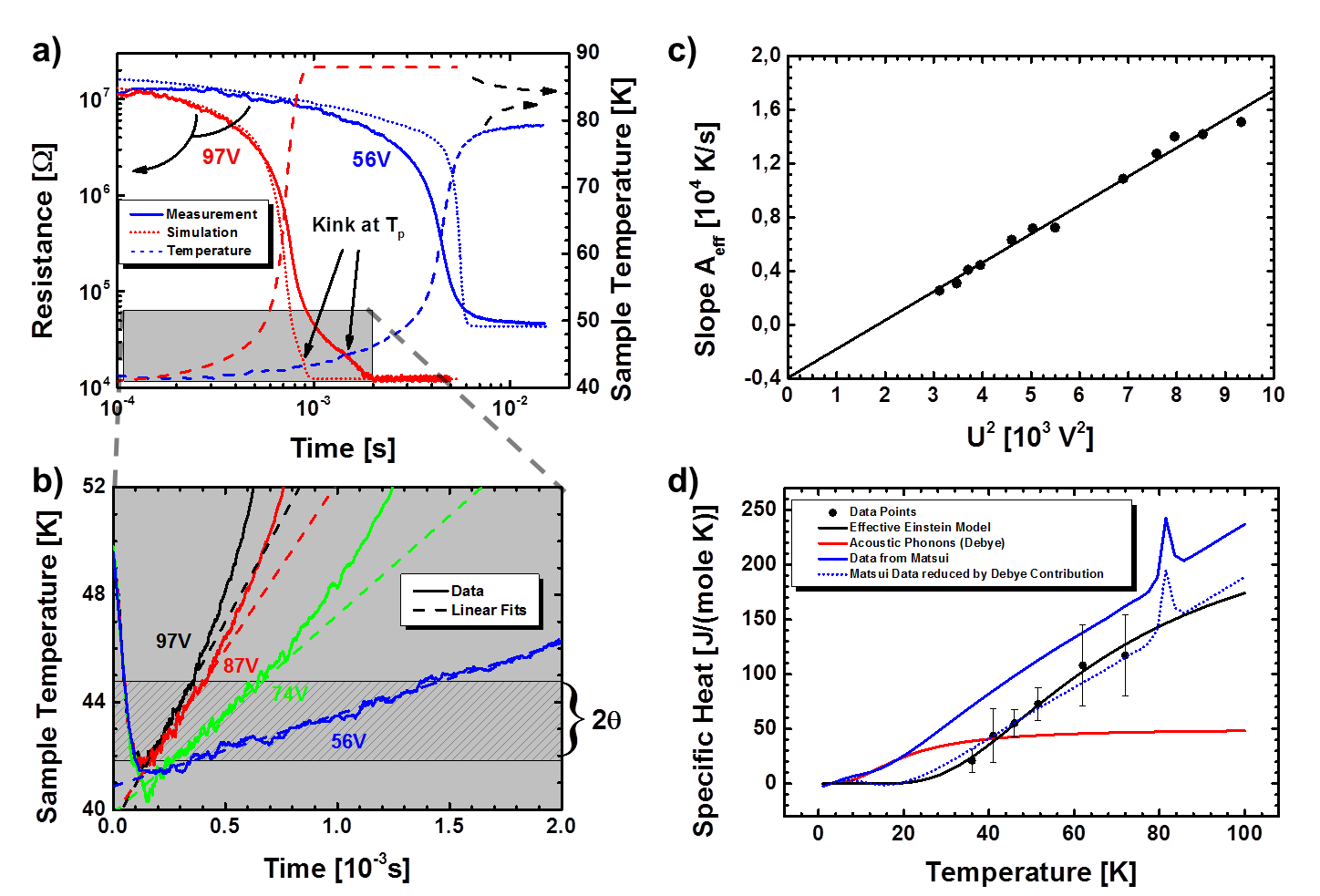}
	\caption{ (a) Measured (solid line) and simulated (dotted line) dynamic resistance response to voltage pulses of 56 V and 97 V and sample temperature (dashed line) calculated from the measured resistance at $T_0=$ 40 K. At high voltages a kink is present in the measurement as well as in the simulations indicating when the phase transition temperature is reached. (b) Zoom-in on the initial temperature rise together with linear fits for different voltages. The slope $A_{eff}$ of the linear fit reveals a (c) $U^2$-dependence from which the (d) specific heat can be determined for each temperature measured. In addition to the literature data by Matsui \cite{Matsui2009} and the expected acoustic phonon contribution, the fit of the effective Einstein model to our data is displayed. In analogy to the approach reported Matsui et al. we also added the latent heat of the phase transition (dashed line) to the Einstein model.}
\label{fig:DetSpecHeat}	
\end{figure*}

The measured dynamic response of the sample resistance at $T_0=$ 40 K to voltage pulses of $U=$ 56 V and $U=$ 97 V are exemplary illustrated in Fig. \ref{fig:DetSpecHeat}a together with simulations assuming fixed values of $c_m$ and  $\widetilde{\alpha}$. After a time-delay on the order of milliseconds, the resistance of the sample rapidly changes from a high-resistive to a low-resistive state. At high voltages a kink can be seen in the measurement data as well as in the simulations. It is caused by the huge drop of the resistivity at the phase transition and indicates when the transition temperature is reached, i.e. the charge order is melted. Adjusting the steady-state values of  measurement and simulation after switching we determined $\widetilde{\alpha}=(9 \pm 1)\cdot 10^{-5}$ W/K in agreement with the value obtained in Sec. \ref{sec:res-CV}. The transient behavior is only fairly reproduced by the simulation, especially when the increase in sample temperature becomes large, i.e. on intermediate time-scales. This stems from an inaccurate determination of $c_m=(55 \pm 25)$ J/(mole K) and its temperature dependence, the latter not being considered in the simulations. However, from the magnitude we can infer that the nonlinear conduction in (DCNQI-d$_6$)$_2$Cu is thermally driven, thus making it reasonable to treat the temperature parameter in the electrothermal model as the sample temperature.

In order to estimate the specific heat and its temperature dependence more accurately and directly from our measurement data, we adopted the following evaluation procedure known from the transient electrothermal method \cite{Guo2007}. First of all, we converted the transient resistance into a transient sample temperature with the help of the temperature-dependent resistance curve measured and displayed in Fig. \ref{fig:Nonlinear-CV-Characteristics}a. The resulting curves are also illustrated in Fig. \ref{fig:DetSpecHeat}a. Zooming-in the initial temperature rise we found a linear relationship of the sample temperature on short time-scales (see Fig. \ref{fig:DetSpecHeat}b) as anticipated by Eq. \ref{eq:electrothermal-model-lowT}. The shaded area in Fig. \ref{fig:DetSpecHeat}b corresponds to the overall temperature increase $2\theta$ in the fitting regime. The expected $U^2$-dependence of the slope $A_{eff}$ in the linear regime is depicted in Fig. \ref{fig:DetSpecHeat}c. With Eq. \ref{eq:electrothermal-model-lowT} we can determine the molar specific heat from the slope $B=\Delta A_{eff}^2/\Delta U^2$ by:

\begin{equation}
	c_m=\frac{m_{mol}}{\rho V B R_{S,eff} \left\{1+R_L/R_{S,eff}\right\}^2}
	\label{eq:MolSpecHeat}
\end{equation}

Here, $R_{S,eff}=R_S(T_0+\theta)$ represents the average resistance of the sample during the initial temperature rise $2\Theta$ linearly fitted with an error of $\sigma_R=1/2\cdot(R_S[T_0+2\theta]-R_S[T_0])$. According to Eq. \ref{eq:electrothermal-model-lowT}, we additionally estimated $\widetilde{\alpha}=(7 \pm 4)\cdot 10^{-5}$ W/K from the intercept C of the linear fits of $A_{eff}$ for various temperatures. Given the low sensitivity of the initial temperature rise on $\widetilde{\alpha}$, this value is in good agreement with the previously determined values of $\widetilde{\alpha}=9\cdot 10^{-5}$ W/K. \\
Including the error of $R_{S,eff}$ and the slope B, the molar specific heat $c_m$ determined at different temperatures is presented in Fig. \ref{fig:DetSpecHeat}d. The qualitative temperature dependence and the magnitude of $c_m$ agree quite well with specific heat data of high accuracy published by Matsui \cite{Matsui2009}. Nonetheless, within the error our measurement systematically undermines the published data by a rather constant offset of about 40 J/(mole K). Even an inaccuracy in the optical determination of the sample volume cannot account for this  deviation. 
In general, the specific heat can be assumed to consist of contributions from acoustic and optical phonons as well as electrons:

\begin{equation}
	c_m=C_{ac}+C_{opt}+C_{el}\\
	\label{eq:SpecHeatContributions}\\
\end{equation}

The linear temperature dependence of our specific heat data (not shown) with a slope of $(3.0 \pm 0.2)$ J/(mole K$^2$) may be considered to originate from the electronic part $C_{el}=\gamma \cdot T$. Yet, the electronic specific heat coefficient $\gamma=0.025$ J/(mole K$^2$) \cite{Nishio2000} in the metallic state of DCNQI$_2$Cu is two orders of magnitude smaller and therefore, should be negligible in the investigated insulating phase. Consequently, we disregard this contribution (and any other from low-energy charge-carrying excitations\cite{Lasjaunias1996}). In solids states with three translational and three rotational degrees of freedom, the specific heat contribution from acoustic phonons can be calculated by the Debye model\cite{Sallamie2005}:

\begin{equation}
	C_{ac}=18 R \left(\frac{T}{\Theta_D}\right)^3 \int_0^{\Theta_D/T}\frac{x^4 e^{x}}{(e^{x}-1)^2}
	\label{eq:Debye-Model}
\end{equation}

$R$ is the ideal gas constant and $\Theta_D$ the Debye temperature. Assuming a Debye temperature of $\Theta_D=$ 82 K\cite{Nishio2000}, this contribution to the specific heat is also depicted in Fig. \ref{fig:DetSpecHeat}d. According to the six degrees of freedom taken into account, the Debye contribution by acoustic phonons approaches a high temperature limit of 6$R\approx$ 50 J/(mole K). In spite of being on the same order of magnitude as our data, acoustic phonons only account for up to about 44 J/(mole K) in the investigated temperature regime. Furthermore, the variation between 35 K and 75 K is quite small due to the low Debye temperature of the material. Together with the observation that our specific heat data is systematically lower compared to the published data by Matsui et al. this suggests, that the saturated contribution of  low-energy acoustic phonons to the specific heat does not allow for transient heating of our sample by inelastic scattering of charge carriers. Additional degrees of freedom are available by optical lattice phonons (external modes) and intramolecular vibrations (internal modes). Taking into account $N=$ 22 atoms per molecule, there are $3N-6=$ 60 internal modes and $6Z-6=$ 30 external modes, $Z=$ 6 denoting the number of building blocks constituting the unit cell of (DCNQI-d$_6$)$_2$Cu single crystals. We estimated the contribution from optical phonons to the specific heat in (DCNQI-d$_6$)$_2$Cu by subtracting the Debye heat from the Matsui data and found a good agreement with our measurement data. Due to the low dispersion of both internal and external modes, their contribution to the specific heat in molecular solids is reasonably well described by a series of harmonic oscillators\cite{Sallamie2005}. To estimate the mean energy of the contributing optical phonon modes, we approximate our data by an effective Einstein model \cite{Wei1979}

\begin{equation}
	C_{opt}=N_E R \left(\frac{\Theta_E}{T}\right)^2 \frac{e^{\Theta_E/T}}{(e^{\Theta_E/T}-1)^2}
	\label{eq:simple-Einstein-Model}
\end{equation}

where $N_E$ is the number of oscillators with an effective frequency $\nu_E=k_B \Theta_E/h$, and $\Theta_E$ being the Einstein temperature. We obtained the best results by neglecting the electronic ($C_{el}$) and acoustic phonon ($C_{ac}$) contribution to the specific heat and only fitting the effective Einstein model to our data, i.e. only taking into account internal and external optical phonon modes. The fit yields $N_E=30 \pm 3$ and an effective Einstein temperature of $\Theta_E=(211\pm 9)$ K corresponding to an average energy of the contributing phonon modes of $\overline{E_{ph}}=19$ meV. This mean energy correlates very well with the weighted optical phonon energies between 7-24 meV observed by Raman spectroscopy in DCNQI$_2$Cu\cite{Sekine1996}. Although the effective number of oscillators equals the number of possible external modes, it is most likely that internal and mixed modes contribute as well\cite{Dlott1986}. Hence, we conclude that intra- and intermolecular optical modes govern the temperature dependence of the effective specific heat in the investigated regime. To accurately simulate the complete dynamic resistance response we used the effective Einstein model of the specific heat. The resulting transient simulations will be presented in the following.\\

\subsection{Simulation Dynamic Resistive Switching}
\label{sec:Res-SimuDRS}

\begin{figure*}
	\includegraphics[width=1\textwidth]{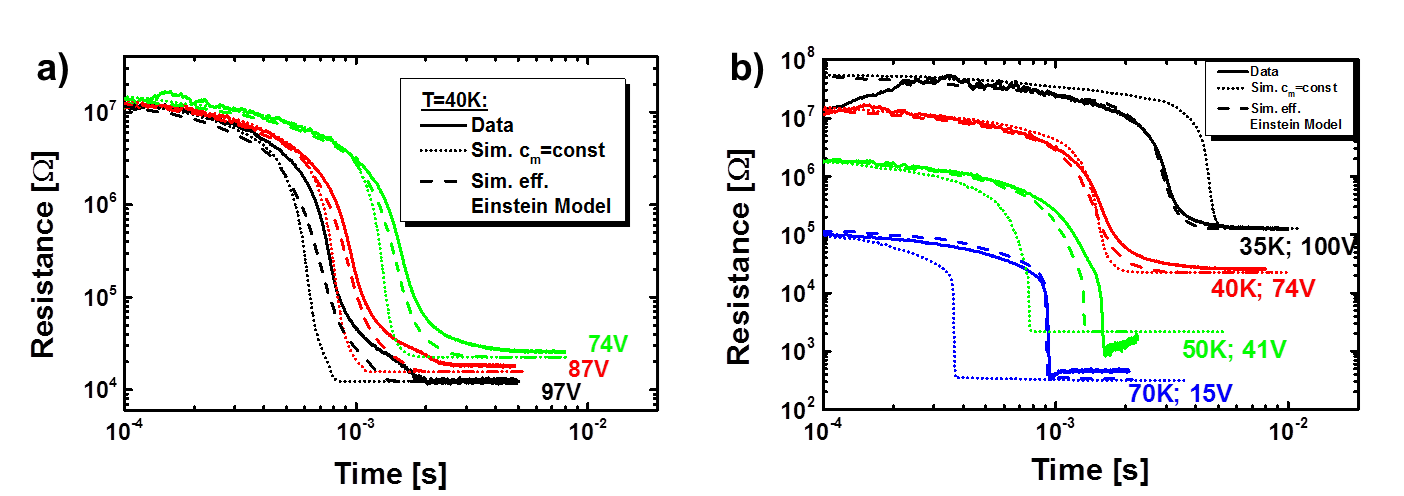}
	\caption{Comparison of measured (solid lines) and simulated dynamic resistance response to (a) different voltage pulses at $T_0=40K$ and (b) at different temperatures utilizing a constant specific heat of $c_m=$ 55 J/(mole K) (dotted lines) and the effective Einstein model of the specific heat (dashed lines) as presented in Sec \ref{sec:Res-DR}.}
\label{fig:Transients}	
\end{figure*}

Because of the temperature raise by several ten Kelvin within the samples in response to the large voltage pulses applied we now explicitly take into account the temperature-dependence of $c_m$ in the simulation (Eq. \ref{eq:electrothermal-model-adj}) by means of our Einstein model described in Sec. \ref{sec:Res-DR}. We also added a latent heat contribution (dashed line in Fig. \ref{fig:DetSpecHeat}d) at the phase transition to our model, but found its effect on the transients negligible in the temperature range under study as it only contributes at large temperature differences between sample and environment, i.e. where the K-term already dominates Eq. \ref{eq:electrothermal-model-adj}. Fig. \ref{fig:Transients}a shows the transient resistance at $T_0=$ 40 K in response to voltage pulses of different magnitude together with simulations, both assuming a constant specific heat of $c_m=$ 55 J/(mole K) (dotted line) and employing the temperature-dependent Einstein Model (dashed line). The transients are far better reproduced including explicitly the temperature-dependence of $c_m$ in the simulations. In particular, implementing the effective Einstein Model we can reproduce the transients at various temperatures to a much better extent compared to the case of a constant specific heat (Fig. \ref{fig:Transients}b). The small dip in the resistance curves after switching for temperatures above $50$ K is due to the current limiter of the pulsed voltage source which was used to prevent sample destruction. The overall good agreement between simulations and measurements underlines the accuracy of the specific heat determined according to the procedure described in Sec. \ref{sec:Res-DR}.

\section{Discussion}
\label{sec:Discussion}

We ascribe the deviation of our data from literature to the respective measurement principle utilized. First of all, instead of steady-state conditioned heat flow into and out of the sample\cite{Matsui2009}, we measure the transient non-equilibrium internal heating of a sample caused by current flow at intermediate to large electric fields. While we expect the temperature distribution inside the sample to be of minor relevance according to the short time scales being used to determine $c_m$, a general inhomogeneous current or field distribution inside the crystal might lead to an overestimation of the active charge transport volume in our analysis. The possible formation of an inhomogeneous high conduction state was already discussed by Mori and coworkers\cite{Mori2009}. This interpretation is supported by flicker noise studies revealing a pronounced noise level enhancement in organic charge-transfer salts which was also ascribed to the presence of an inhomogeneous current flow\cite{Galchenkov1998, Mueller2009}. Additionally, we were restricted to two-probe measurements due to the high fields necessary to induce the resistive switching. The non-planar silver paint contacts therefore might facilitate an inhomogeneous field distribution accompanied by local fluctuations of the charge carrier density.\\

Nevertheless, being able to accurately model the specific heat data by an effective Einstein model solely based on optical phonon modes, we rather explain the experimental findings by extending the electrothermal model as shown in Fig. \ref{fig:model} and allowing for an efficient stimulation of optical phonons by interacting with charge-carrying excitations. In contrast to acoustic phonons, the energy transferred to the optical phonon system remains within the sample as the contribution of these lattice excitations to the thermal conductivity is small due to their low group velocity. This means that the population density of the optical phonon modes is shifted out of equilibrium and its temperature decouples from that of the acoustic phonon system. This excess energy facilitates the generation of additional charge carrier excitations leading to a multiplication effect. The mechanism of nonlinear conduction originating from current induced non-equilibrium optical phonon population has already been demonstrated in quasi-1D single-walled carbon nanotubes\cite{Pop2005} as well as in 2D graphene\cite{Berciaud2010}.\\
The energy to excite additional charge carriers is estimated from the Boltzmann-like activation of the DC conductivity with an energy of about $E_{a}=(39\pm5)$ meV slightly varying with temperature across the investigated regime (see Fig. \ref{fig:Nonlinear-CV-Characteristics}a). This energy is considerably smaller than the expected Peierls gap of $E_{Gap}\approx 100$ meV and has been ascribed to an excitation of charge carriers from shallow states within the band gap\cite{Bauer1993}. The activation energy related to electronic transport is about twice as large as the observed effective phonon energy of $\overline{E_{ph}}=19$ meV, for which reason two-phonon processes might be involved in the excitation of charge-carriers in the CDW state of DCNQI$_2$Cu. In general, an efficient interaction between charge-carrying excitations and optical phonons seems to be essential for the observed conductivity phenomena. \\

Single excited charge carriers would need to gain the energy of $\overline{E_{ph}}=19$ meV by the electric field to scatter with these optical phonons. Assuming an effective mass of $m_{eff}=3.35m_0$ \cite{Hill1996} and an applied electric field of about $E\leq 3600$ V/cm, we estimate the necessary mobility of charge carriers to be $\mu \geq \sqrt{2 \overline{E_{ph}}/m_{eff}}/E=1200$ cm$^2$/(Vs). This value is quite high but not unrealistic considering Hall mobilities on the order of $10^5$ cm$^2$/(Vs) which have been measured at 4 K for one-dimensional organic conductors such as (TMTSF)$_2$PF$_6$\cite{Chaikin1981}. Additionally, in the organic semiconductor naphthalene the saturation of charge carrier velocity at cryogenic temperatures and high electric fields has also been ascribed to optical phonon generation via inelastic scattering\cite{Karl2003}. Alternatively, low-energy excitations of the electronic system might be constituted by soliton-like charged domain walls, which have been proposed to explain the low-frequency dielectric response in the CDW state of DCNQI$_2$Cu\cite{pinterić2000low}. Soliton-like excitations are known to play a crucial role in the transport of commensurate CDW systems like polyacetylene \cite{Su1980} as well as of various organic charge-transfer salts, e.g. $\alpha$-(BEDT-TTF)$_2$I$_3$ \cite{Ivek2010}. It was noted that such solitons are expected to couple more efficiently to optical than to acoustic phonon modes by a factor of $\xi/a$ where $a$ denotes the molecular lattice spacing along the chain direction and $\xi$ is the spatial extension of the soliton\cite{Maki1982}. Thus, a large spatial extension of the soliton with respect to the lattice constant might lead to preferred interaction with optical phonons as indicated by our determined effective specific heat. While we cannot completely resolve the fundamental conduction mechanism from our data, an efficient interaction between the optical phonon modes and the charge-carrying excitations seems to be essential for the occurrence of nonlinear conduction effects in DCNQI$_2$Cu.\\
For the sake of completeness, we also estimated the main loss mechanism of thermal energy to the environment. If the energy would be transferred from the crystal surface to the surroundings via convection or radiation, our determined $\widetilde{\alpha}=10^{-5}$ W/K would yield a heat transfer coefficient of $h=\widetilde{\alpha}/A_S=3000$ W/(m$^2$K) with $A_S=3\cdot10^{-4}$ cm$^2$ being the surface area of the sample. This heat transfer coefficient is much larger than that expected for convection ($h_{conv}=6-30$ W/(m$^2$ K))\cite{Kreith2010}. With the Stefan-Boltzmann constant $\sigma_S$, a maximum crystal surface temperature of $T_1=$ 90 K and a minimum ambient temperature of $T_0=$ 35 K, we can estimate the coefficient of radiative heat transfer by $h_{rad}=\sigma_S\cdot(T_1+T_0)(T_1^2+T_0^2)=0.07$ W/(m$^2$ K) which, according to its magnitude, is also neglected\cite{Kreith2010}. The conductive heat transfer to the contacts is determined by the thermal conductivity $\kappa$ of the material which we roughly estimate by $\kappa \leq \widetilde{\alpha}\cdot l/(8\cdot A_{cs})$, $l$ being the length and $A_{cs}$ the cross-section area of the sample\cite{Brill1981}. We obtain a value of $\kappa \leq 3.4$ W/(m K) which agrees reasonably well with the thermal conductivity of about 1.3 W/(m K) determined for alloyed DCNQI$_2$Li$_{0.86}$Cu$_{0.14}$ at 50 K, bearing in mind the increase of thermal conductivity with increasing copper content\cite{Torizuka2005a}. Together with the assignment to acoustic phonons, the absence of appreciable variations of this thermal conductivity between 30 K and 150 K confirms the interpretation of our temperature-independent $\widetilde{\alpha}$ value. Hence, we identify the main energy loss mechanism in our samples to be the energy transferred to the acoustic phonon system which is efficiently transported to the contacts via thermal conduction as depicted in our model (see Fig. \ref{fig:model}). This energy exchange to acoustic phonons can occur via charge carrier-phonon scattering ($\alpha_{e-ac}$) or via scattering between optical and acoustic phonons ($\alpha_{opt-ac}$). Both mechanisms seem to be inefficient at high electric fields supporting the occurrence of the measured nonlinear conduction effects.\\

\section{Conclusion}

We experimentally studied the nonlinear conductivity in (DCNQI-d$_6$)$_2$Cu single crystals by analyzing the dynamical resistance in response to large voltage pulses. In extension to simulations of the current-voltage characteristics at fixed pulse lengths, we demonstrate that applying the electrothermal description of low-dimensional conductors\cite{Mori2009} to the entire transient resistivity provides a more detailed insight into the energy exchange between the phononic and electronic subsystems of the material. Our simulations are able to accurately reproduce the nonlinear conduction behavior of our samples and reveal a distinct non-equilibrium excitation of optical phonon modes at a mean energy of about 19 meV. We infer a pronounced coupling between the electronic and optical phonon degrees of freedom at high electric fields in this material class. Due to their small contribution to the thermal conductivity, the energy transferred to the optical phonon system remains within the sample and can recover by additional single or collective charge carrying excitations leading to a thermally induced multiplication effect. Energy channeled into the acoustic phonon system is efficiently transferred to the contacts via heat conduction. As the involved effective optical phonon energy of around 19 meV is about half the activation energy of 39 meV observed in the DC conductivity, an excitation process including two phonons can be anticipated. Therefore, the nonlinear conduction in (DCNQI-d$_6$)$_2$Cu crucially depends on the charge carrier interaction with optical phonon modes relative to the excitation of acoustic phonons via interaction of electrons or optical phonons with acoustic phonons. Its origin seems to be an interplay of population density of low-energy optical phonon states, small transport activation energy, large mobility at low temperatures and strong coupling between optical phonon modes and charge carriers at high fields in this material class. Hence, our results bridge the gap between the electrothermal model phenomenologically introduced by Mori et al. and the microscopic origin of nonlinear transport effects in DCNQI$_2$Cu. As such, we consider the proposed mechanism also valid for nonlinear conduction phenomena in other low-dimensional molecular conductors with different ground states.

\begin{acknowledgments}
Financial support by the DFG (project PF385/6-1) and the 7th Framework Program (H2ESOT) of the European Commission are gratefully acknowledged. We like to thank Markus Schwoerer (University of Bayreuth) and Martin Dressel (University of Stuttgart) for fruitful discussions and helpful comments.
\end{acknowledgments}

\bibliography{library}

\begin{thebibliography}{36}%
\makeatletter
\providecommand \@ifxundefined [1]{%
 \@ifx{#1\undefined}
}%
\providecommand \@ifnum [1]{%
 \ifnum #1\expandafter \@firstoftwo
 \else \expandafter \@secondoftwo
 \fi
}%
\providecommand \@ifx [1]{%
 \ifx #1\expandafter \@firstoftwo
 \else \expandafter \@secondoftwo
 \fi
}%
\providecommand \natexlab [1]{#1}%
\providecommand \enquote  [1]{``#1''}%
\providecommand \bibnamefont  [1]{#1}%
\providecommand \bibfnamefont [1]{#1}%
\providecommand \citenamefont [1]{#1}%
\providecommand \href@noop [0]{\@secondoftwo}%
\providecommand \href [0]{\begingroup \@sanitize@url \@href}%
\providecommand \@href[1]{\@@startlink{#1}\@@href}%
\providecommand \@@href[1]{\endgroup#1\@@endlink}%
\providecommand \@sanitize@url [0]{\catcode `\\12\catcode `\$12\catcode
  `\&12\catcode `\#12\catcode `\^12\catcode `\_12\catcode `\%12\relax}%
\providecommand \@@startlink[1]{}%
\providecommand \@@endlink[0]{}%
\providecommand \url  [0]{\begingroup\@sanitize@url \@url }%
\providecommand \@url [1]{\endgroup\@href {#1}{\urlprefix }}%
\providecommand \urlprefix  [0]{URL }%
\providecommand \Eprint [0]{\href }%
\providecommand \doibase [0]{http://dx.doi.org/}%
\providecommand \selectlanguage [0]{\@gobble}%
\providecommand \bibinfo  [0]{\@secondoftwo}%
\providecommand \bibfield  [0]{\@secondoftwo}%
\providecommand \translation [1]{[#1]}%
\providecommand \BibitemOpen [0]{}%
\providecommand \bibitemStop [0]{}%
\providecommand \bibitemNoStop [0]{.\EOS\space}%
\providecommand \EOS [0]{\spacefactor3000\relax}%
\providecommand \BibitemShut  [1]{\csname bibitem#1\endcsname}%
\let\auto@bib@innerbib\@empty
\bibitem [{\citenamefont {Iwasa}\ \emph {et~al.}(1989)\citenamefont {Iwasa},
  \citenamefont {Koda}, \citenamefont {Koshihara}, \citenamefont {Tokura},
  \citenamefont {Iwasawa},\ and\ \citenamefont {Saito}}]{Iwasa1989}%
  \BibitemOpen
  \bibfield  {author} {\bibinfo {author} {\bibfnamefont {Y.}~\bibnamefont
  {Iwasa}}, \bibinfo {author} {\bibfnamefont {T.}~\bibnamefont {Koda}},
  \bibinfo {author} {\bibfnamefont {S.}~\bibnamefont {Koshihara}}, \bibinfo
  {author} {\bibfnamefont {Y.}~\bibnamefont {Tokura}}, \bibinfo {author}
  {\bibfnamefont {N.}~\bibnamefont {Iwasawa}}, \ and\ \bibinfo {author}
  {\bibfnamefont {G.}~\bibnamefont {Saito}},\ }\href {\doibase
  10.1103/PhysRevB.39.10441} {\bibfield  {journal} {\bibinfo  {journal}
  {Physical Review B}\ }\textbf {\bibinfo {volume} {39}},\ \bibinfo {pages}
  {10441} (\bibinfo {year} {1989})}\BibitemShut {NoStop}%
\bibitem [{\citenamefont {Mori}\ \emph {et~al.}(2008)\citenamefont {Mori},
  \citenamefont {Bando}, \citenamefont {Kawamoto}, \citenamefont {Terasaki},
  \citenamefont {Takimiya},\ and\ \citenamefont {Otsubo}}]{Mori2008}%
  \BibitemOpen
  \bibfield  {author} {\bibinfo {author} {\bibfnamefont {T.}~\bibnamefont
  {Mori}}, \bibinfo {author} {\bibfnamefont {Y.}~\bibnamefont {Bando}},
  \bibinfo {author} {\bibfnamefont {T.}~\bibnamefont {Kawamoto}}, \bibinfo
  {author} {\bibfnamefont {I.}~\bibnamefont {Terasaki}}, \bibinfo {author}
  {\bibfnamefont {K.}~\bibnamefont {Takimiya}}, \ and\ \bibinfo {author}
  {\bibfnamefont {T.}~\bibnamefont {Otsubo}},\ }\href {\doibase
  10.1103/PhysRevLett.100.037001} {\bibfield  {journal} {\bibinfo  {journal}
  {Physical Review Letters}\ }\textbf {\bibinfo {volume} {100}},\ \bibinfo
  {pages} {037001} (\bibinfo {year} {2008})}\BibitemShut {NoStop}%
\bibitem [{\citenamefont {Kumai}\ \emph {et~al.}(1999)\citenamefont {Kumai},
  \citenamefont {Okimoto},\ and\ \citenamefont {Tokura}}]{Kumai1999}%
  \BibitemOpen
  \bibfield  {author} {\bibinfo {author} {\bibfnamefont {R.}~\bibnamefont
  {Kumai}}, \bibinfo {author} {\bibfnamefont {Y.}~\bibnamefont {Okimoto}}, \
  and\ \bibinfo {author} {\bibfnamefont {Y.}~\bibnamefont {Tokura}},\ }\href
  {http://www.sciencemag.org/content/284/5420/1645.abstract} {\bibfield
  {journal} {\bibinfo  {journal} {Science}\ }\textbf {\bibinfo {volume}
  {284}},\ \bibinfo {pages} {1645} (\bibinfo {year} {1999})}\BibitemShut
  {NoStop}%
\bibitem [{\citenamefont {Tamura}\ \emph {et~al.}(2010)\citenamefont {Tamura},
  \citenamefont {Ozawa}, \citenamefont {Bando}, \citenamefont {Kawamoto},\ and\
  \citenamefont {Mori}}]{Tamura2010}%
  \BibitemOpen
  \bibfield  {author} {\bibinfo {author} {\bibfnamefont {K.}~\bibnamefont
  {Tamura}}, \bibinfo {author} {\bibfnamefont {T.}~\bibnamefont {Ozawa}},
  \bibinfo {author} {\bibfnamefont {Y.}~\bibnamefont {Bando}}, \bibinfo
  {author} {\bibfnamefont {T.}~\bibnamefont {Kawamoto}}, \ and\ \bibinfo
  {author} {\bibfnamefont {T.}~\bibnamefont {Mori}},\ }\href {\doibase
  10.1063/1.3428388} {\bibfield  {journal} {\bibinfo  {journal} {Journal of
  Applied Physics}\ }\textbf {\bibinfo {volume} {107}},\ \bibinfo {pages}
  {103716} (\bibinfo {year} {2010})}\BibitemShut {NoStop}%
\bibitem [{\citenamefont {Ivek}\ \emph {et~al.}(2012)\citenamefont {Ivek},
  \citenamefont {Kova\v{c}evi\'{c}}, \citenamefont {Pinteri\'{c}},
  \citenamefont {Korin-Hamzi\'{c}}, \citenamefont {Tomi\'{c}}, \citenamefont
  {Knoblauch}, \citenamefont {Schweitzer},\ and\ \citenamefont
  {Dressel}}]{Ivek2012}%
  \BibitemOpen
  \bibfield  {author} {\bibinfo {author} {\bibfnamefont {T.}~\bibnamefont
  {Ivek}}, \bibinfo {author} {\bibfnamefont {I.}~\bibnamefont
  {Kova\v{c}evi\'{c}}}, \bibinfo {author} {\bibfnamefont {M.}~\bibnamefont
  {Pinteri\'{c}}}, \bibinfo {author} {\bibfnamefont {B.}~\bibnamefont
  {Korin-Hamzi\'{c}}}, \bibinfo {author} {\bibfnamefont {S.}~\bibnamefont
  {Tomi\'{c}}}, \bibinfo {author} {\bibfnamefont {T.}~\bibnamefont
  {Knoblauch}}, \bibinfo {author} {\bibfnamefont {D.}~\bibnamefont
  {Schweitzer}}, \ and\ \bibinfo {author} {\bibfnamefont {M.}~\bibnamefont
  {Dressel}},\ }\href {\doibase 10.1103/PhysRevB.86.245125} {\bibfield
  {journal} {\bibinfo  {journal} {Physical Review B}\ }\textbf {\bibinfo
  {volume} {86}},\ \bibinfo {pages} {245125} (\bibinfo {year}
  {2012})}\BibitemShut {NoStop}%
\bibitem [{\citenamefont {Mori}\ \emph {et~al.}(2009)\citenamefont {Mori},
  \citenamefont {Ozawa}, \citenamefont {Bando}, \citenamefont {Kawamoto},
  \citenamefont {Niizeki}, \citenamefont {Mori},\ and\ \citenamefont
  {Terasaki}}]{Mori2009}%
  \BibitemOpen
  \bibfield  {author} {\bibinfo {author} {\bibfnamefont {T.}~\bibnamefont
  {Mori}}, \bibinfo {author} {\bibfnamefont {T.}~\bibnamefont {Ozawa}},
  \bibinfo {author} {\bibfnamefont {Y.}~\bibnamefont {Bando}}, \bibinfo
  {author} {\bibfnamefont {T.}~\bibnamefont {Kawamoto}}, \bibinfo {author}
  {\bibfnamefont {S.}~\bibnamefont {Niizeki}}, \bibinfo {author} {\bibfnamefont
  {H.}~\bibnamefont {Mori}}, \ and\ \bibinfo {author} {\bibfnamefont
  {I.}~\bibnamefont {Terasaki}},\ }\href {\doibase 10.1103/PhysRevB.79.115108}
  {\bibfield  {journal} {\bibinfo  {journal} {Physical Review B}\ }\textbf
  {\bibinfo {volume} {79}},\ \bibinfo {pages} {115108} (\bibinfo {year}
  {2009})}\BibitemShut {NoStop}%
\bibitem [{\citenamefont {Aum\"{u}ller}\ \emph {et~al.}(1986)\citenamefont
  {Aum\"{u}ller}, \citenamefont {Erk}, \citenamefont {Klebe}, \citenamefont
  {H\"{u}nig}, \citenamefont {von Sch\"{u}tz},\ and\ \citenamefont
  {Werner}}]{Aumueller1986}%
  \BibitemOpen
  \bibfield  {author} {\bibinfo {author} {\bibfnamefont {A.}~\bibnamefont
  {Aum\"{u}ller}}, \bibinfo {author} {\bibfnamefont {P.}~\bibnamefont {Erk}},
  \bibinfo {author} {\bibfnamefont {G.}~\bibnamefont {Klebe}}, \bibinfo
  {author} {\bibfnamefont {S.}~\bibnamefont {H\"{u}nig}}, \bibinfo {author}
  {\bibfnamefont {J.}~\bibnamefont {von Sch\"{u}tz}}, \ and\ \bibinfo {author}
  {\bibfnamefont {H.}~\bibnamefont {Werner}},\ }\href
  {http://onlinelibrary.wiley.com/doi/10.1002/anie.198607401/abstract}
  {\bibfield  {journal} {\bibinfo  {journal} {Angewandte Chemie International
  Edition in English}\ }\textbf {\bibinfo {volume} {25}},\ \bibinfo {pages}
  {740} (\bibinfo {year} {1986})}\BibitemShut {NoStop}%
\bibitem [{\citenamefont {Kato}\ \emph {et~al.}(1993)\citenamefont {Kato},
  \citenamefont {Sawa}, \citenamefont {Aonuma}, \citenamefont {Tamura},
  \citenamefont {Kinoshita},\ and\ \citenamefont {Kobayashi}}]{Kato1993}%
  \BibitemOpen
  \bibfield  {author} {\bibinfo {author} {\bibfnamefont {R.}~\bibnamefont
  {Kato}}, \bibinfo {author} {\bibfnamefont {H.}~\bibnamefont {Sawa}}, \bibinfo
  {author} {\bibfnamefont {S.}~\bibnamefont {Aonuma}}, \bibinfo {author}
  {\bibfnamefont {M.}~\bibnamefont {Tamura}}, \bibinfo {author} {\bibfnamefont
  {M.}~\bibnamefont {Kinoshita}}, \ and\ \bibinfo {author} {\bibfnamefont
  {H.}~\bibnamefont {Kobayashi}},\ }\href {\doibase
  10.1016/0038-1098(93)90187-R} {\bibfield  {journal} {\bibinfo  {journal}
  {Solid State Communications}\ }\textbf {\bibinfo {volume} {85}},\ \bibinfo
  {pages} {831} (\bibinfo {year} {1993})}\BibitemShut {NoStop}%
\bibitem [{\citenamefont {Wakita}\ \emph {et~al.}(2010)\citenamefont {Wakita},
  \citenamefont {Ozawa}, \citenamefont {Bando},\ and\ \citenamefont
  {Mori}}]{Wakita2010}%
  \BibitemOpen
  \bibfield  {author} {\bibinfo {author} {\bibfnamefont {H.}~\bibnamefont
  {Wakita}}, \bibinfo {author} {\bibfnamefont {T.}~\bibnamefont {Ozawa}},
  \bibinfo {author} {\bibfnamefont {Y.}~\bibnamefont {Bando}}, \ and\ \bibinfo
  {author} {\bibfnamefont {T.}~\bibnamefont {Mori}},\ }\href {\doibase
  10.1143/JPSJ.79.094703} {\bibfield  {journal} {\bibinfo  {journal} {Journal
  of the Physical Society of Japan}\ }\textbf {\bibinfo {volume} {79}},\
  \bibinfo {pages} {094703} (\bibinfo {year} {2010})}\BibitemShut {NoStop}%
\bibitem [{\citenamefont {Vuleti\'{c}}\ \emph {et~al.}(2001)\citenamefont
  {Vuleti\'{c}}, \citenamefont {Pinteri\'{c}}, \citenamefont
  {Lon\v{c}ari\'{c}}, \citenamefont {Tomi\'{c}},\ and\ \citenamefont {von
  Sch\"{u}tz}}]{Vuletic2001}%
  \BibitemOpen
  \bibfield  {author} {\bibinfo {author} {\bibfnamefont {T.}~\bibnamefont
  {Vuleti\'{c}}}, \bibinfo {author} {\bibfnamefont {M.}~\bibnamefont
  {Pinteri\'{c}}}, \bibinfo {author} {\bibfnamefont {M.}~\bibnamefont
  {Lon\v{c}ari\'{c}}}, \bibinfo {author} {\bibfnamefont {S.}~\bibnamefont
  {Tomi\'{c}}}, \ and\ \bibinfo {author} {\bibfnamefont {J.~J.}\ \bibnamefont
  {von Sch\"{u}tz}},\ }\href {\doibase 10.1016/S0379-6779(00)00933-4}
  {\bibfield  {journal} {\bibinfo  {journal} {Synthetic Metals}\ }\textbf
  {\bibinfo {volume} {120}},\ \bibinfo {pages} {1001} (\bibinfo {year}
  {2001})}\BibitemShut {NoStop}%
\bibitem [{\citenamefont {Pinteri\'{c}}\ \emph {et~al.}(2000)\citenamefont
  {Pinteri\'{c}}, \citenamefont {Vuleti\'{c}}, \citenamefont
  {Lon\v{c}ari\'{c}}, \citenamefont {Tomi\'{c}},\ and\ \citenamefont {{Von
  Sch\"{u}tz}}}]{pinterić2000low}%
  \BibitemOpen
  \bibfield  {author} {\bibinfo {author} {\bibfnamefont {M.}~\bibnamefont
  {Pinteri\'{c}}}, \bibinfo {author} {\bibfnamefont {T.}~\bibnamefont
  {Vuleti\'{c}}}, \bibinfo {author} {\bibfnamefont {M.}~\bibnamefont
  {Lon\v{c}ari\'{c}}}, \bibinfo {author} {\bibfnamefont {S.}~\bibnamefont
  {Tomi\'{c}}}, \ and\ \bibinfo {author} {\bibfnamefont {J.~U.}\ \bibnamefont
  {{Von Sch\"{u}tz}}},\ }\href
  {http://www.springerlink.com/index/2J2X3ENJ14RYQ0EL.pdf} {\bibfield
  {journal} {\bibinfo  {journal} {The European Physical Journal B-Condensed
  Matter and Complex Systems}\ }\textbf {\bibinfo {volume} {16}},\ \bibinfo
  {pages} {487} (\bibinfo {year} {2000})}\BibitemShut {NoStop}%
\bibitem [{\citenamefont {Nishio}\ \emph {et~al.}(2000)\citenamefont {Nishio},
  \citenamefont {Tamura}, \citenamefont {Kajita}, \citenamefont {Aonuma},
  \citenamefont {Sawa}, \citenamefont {Kato},\ and\ \citenamefont
  {Kobayashi}}]{Nishio2000}%
  \BibitemOpen
  \bibfield  {author} {\bibinfo {author} {\bibfnamefont {Y.}~\bibnamefont
  {Nishio}}, \bibinfo {author} {\bibfnamefont {M.}~\bibnamefont {Tamura}},
  \bibinfo {author} {\bibfnamefont {K.}~\bibnamefont {Kajita}}, \bibinfo
  {author} {\bibfnamefont {S.}~\bibnamefont {Aonuma}}, \bibinfo {author}
  {\bibfnamefont {H.}~\bibnamefont {Sawa}}, \bibinfo {author} {\bibfnamefont
  {R.}~\bibnamefont {Kato}}, \ and\ \bibinfo {author} {\bibfnamefont
  {H.}~\bibnamefont {Kobayashi}},\ }\href {\doibase 10.1143/JPSJ.69.1414}
  {\bibfield  {journal} {\bibinfo  {journal} {Journal of the Physics Society
  Japan}\ }\textbf {\bibinfo {volume} {69}},\ \bibinfo {pages} {1414} (\bibinfo
  {year} {2000})}\BibitemShut {NoStop}%
\bibitem [{\citenamefont {Matsui}\ \emph {et~al.}(2009)\citenamefont {Matsui},
  \citenamefont {Takaoka}, \citenamefont {Nishio}, \citenamefont {Kato},\ and\
  \citenamefont {Kajita}}]{Matsui2009}%
  \BibitemOpen
  \bibfield  {author} {\bibinfo {author} {\bibfnamefont {A.}~\bibnamefont
  {Matsui}}, \bibinfo {author} {\bibfnamefont {Y.}~\bibnamefont {Takaoka}},
  \bibinfo {author} {\bibfnamefont {Y.}~\bibnamefont {Nishio}}, \bibinfo
  {author} {\bibfnamefont {R.}~\bibnamefont {Kato}}, \ and\ \bibinfo {author}
  {\bibfnamefont {K.}~\bibnamefont {Kajita}},\ }\href {\doibase
  10.1088/1742-6596/150/4/042120} {\bibfield  {journal} {\bibinfo  {journal}
  {Journal of Physics: Conference Series}\ }\textbf {\bibinfo {volume} {150}},\
  \bibinfo {pages} {42120} (\bibinfo {year} {2009})}\BibitemShut {NoStop}%
\bibitem [{\citenamefont {Karl}(2003)}]{Karl2003}%
  \BibitemOpen
  \bibfield  {author} {\bibinfo {author} {\bibfnamefont {N.}~\bibnamefont
  {Karl}},\ }\href {\doibase 10.1016/S0379-6779(02)00398-3} {\bibfield
  {journal} {\bibinfo  {journal} {Synthetic Metals}\ }\textbf {\bibinfo
  {volume} {133-134}},\ \bibinfo {pages} {649} (\bibinfo {year}
  {2003})}\BibitemShut {NoStop}%
\bibitem [{\citenamefont {Sinzger}\ \emph {et~al.}(1993)\citenamefont
  {Sinzger}, \citenamefont {H\"{u}nig}, \citenamefont {Jopp}, \citenamefont
  {Bauer}, \citenamefont {Bietsch}, \citenamefont {von Sch\"{u}tz},
  \citenamefont {Wolf}, \citenamefont {Kremer},\ and\ \citenamefont
  {Metzenthin}}]{Sinzger1993}%
  \BibitemOpen
  \bibfield  {author} {\bibinfo {author} {\bibfnamefont {K.}~\bibnamefont
  {Sinzger}}, \bibinfo {author} {\bibfnamefont {S.}~\bibnamefont {H\"{u}nig}},
  \bibinfo {author} {\bibfnamefont {M.}~\bibnamefont {Jopp}}, \bibinfo {author}
  {\bibfnamefont {D.}~\bibnamefont {Bauer}}, \bibinfo {author} {\bibfnamefont
  {W.}~\bibnamefont {Bietsch}}, \bibinfo {author} {\bibfnamefont {J.~U.}\
  \bibnamefont {von Sch\"{u}tz}}, \bibinfo {author} {\bibfnamefont {H.~C.}\
  \bibnamefont {Wolf}}, \bibinfo {author} {\bibfnamefont {R.~K.}\ \bibnamefont
  {Kremer}}, \ and\ \bibinfo {author} {\bibfnamefont {T.}~\bibnamefont
  {Metzenthin}},\ }\href {\doibase 10.1021/ja00070a013} {\bibfield  {journal}
  {\bibinfo  {journal} {Journal of the American Chemical Society}\ }\textbf
  {\bibinfo {volume} {115}},\ \bibinfo {pages} {7696} (\bibinfo {year}
  {1993})}\BibitemShut {NoStop}%
\bibitem [{\citenamefont {Guo}\ \emph {et~al.}(2007)\citenamefont {Guo},
  \citenamefont {Wang},\ and\ \citenamefont {Wang}}]{Guo2007}%
  \BibitemOpen
  \bibfield  {author} {\bibinfo {author} {\bibfnamefont {J.}~\bibnamefont
  {Guo}}, \bibinfo {author} {\bibfnamefont {X.}~\bibnamefont {Wang}}, \ and\
  \bibinfo {author} {\bibfnamefont {T.}~\bibnamefont {Wang}},\ }\href@noop {}
  {\bibfield  {journal} {\bibinfo  {journal} {Journal of Applied Physics}\
  }\textbf {\bibinfo {volume} {101}} (\bibinfo {year} {2007})}\BibitemShut
  {NoStop}%
\bibitem [{\citenamefont {H\"{u}nig}\ \emph {et~al.}(1998)\citenamefont
  {H\"{u}nig}, \citenamefont {Baub},\ and\ \citenamefont
  {Kemmera}}]{Hunig1998}%
  \BibitemOpen
  \bibfield  {author} {\bibinfo {author} {\bibfnamefont {S.}~\bibnamefont
  {H\"{u}nig}}, \bibinfo {author} {\bibfnamefont {R.}~\bibnamefont {Baub}}, \
  and\ \bibinfo {author} {\bibfnamefont {M.}~\bibnamefont {Kemmera}},\ }\href
  {http://www.erowid.org/archive/rhodium/pdf/2-halo-1.4-dimethoxybenzene.pdf}
  {\bibfield  {journal} {\bibinfo  {journal} {Eur. J. Org. Chem.}\ ,\ \bibinfo
  {pages} {335}} (\bibinfo {year} {1998})}\BibitemShut {NoStop}%
\bibitem [{\citenamefont {Kato}\ \emph {et~al.}(1989)\citenamefont {Kato},
  \citenamefont {Kobayashi},\ and\ \citenamefont {Kobayashi}}]{Kato1989}%
  \BibitemOpen
  \bibfield  {author} {\bibinfo {author} {\bibfnamefont {R.}~\bibnamefont
  {Kato}}, \bibinfo {author} {\bibfnamefont {H.}~\bibnamefont {Kobayashi}}, \
  and\ \bibinfo {author} {\bibfnamefont {A.}~\bibnamefont {Kobayashi}},\ }\href
  {http://pubs.acs.org/doi/abs/10.1021/ja00196a032} {\bibfield  {journal}
  {\bibinfo  {journal} {Journal of the American Chemical Society}\ }\textbf
  {\bibinfo {volume} {111}},\ \bibinfo {pages} {5224} (\bibinfo {year}
  {1989})}\BibitemShut {NoStop}%
\bibitem [{\citenamefont {Bauer}\ \emph {et~al.}(1993)\citenamefont {Bauer},
  \citenamefont {von Sch\"{u}tz}, \citenamefont {Wolf}, \citenamefont
  {H\"{u}nig}, \citenamefont {Sinzger},\ and\ \citenamefont
  {Kremer}}]{Bauer1993}%
  \BibitemOpen
  \bibfield  {author} {\bibinfo {author} {\bibfnamefont {D.}~\bibnamefont
  {Bauer}}, \bibinfo {author} {\bibfnamefont {J.~U.}\ \bibnamefont {von
  Sch\"{u}tz}}, \bibinfo {author} {\bibfnamefont {H.~C.}\ \bibnamefont {Wolf}},
  \bibinfo {author} {\bibfnamefont {S.}~\bibnamefont {H\"{u}nig}}, \bibinfo
  {author} {\bibfnamefont {K.}~\bibnamefont {Sinzger}}, \ and\ \bibinfo
  {author} {\bibfnamefont {R.~K.}\ \bibnamefont {Kremer}},\ }\href
  {http://onlinelibrary.wiley.com/doi/10.1002/adma.19930051109/abstract}
  {\bibfield  {journal} {\bibinfo  {journal} {Advanced Materials}\ }\textbf
  {\bibinfo {volume} {5}},\ \bibinfo {pages} {829} (\bibinfo {year}
  {1993})}\BibitemShut {NoStop}%
\bibitem [{\citenamefont {Lasjaunias}\ \emph {et~al.}(1996)\citenamefont
  {Lasjaunias}, \citenamefont {Biljakovi\'{c}},\ and\ \citenamefont
  {Monceau}}]{Lasjaunias1996}%
  \BibitemOpen
  \bibfield  {author} {\bibinfo {author} {\bibfnamefont {J.~C.}\ \bibnamefont
  {Lasjaunias}}, \bibinfo {author} {\bibfnamefont {K.}~\bibnamefont
  {Biljakovi\'{c}}}, \ and\ \bibinfo {author} {\bibfnamefont {P.}~\bibnamefont
  {Monceau}},\ }\href {\doibase 10.1103/PhysRevB.53.7699} {\bibfield  {journal}
  {\bibinfo  {journal} {Physical Review B}\ }\textbf {\bibinfo {volume} {53}},\
  \bibinfo {pages} {7699} (\bibinfo {year} {1996})}\BibitemShut {NoStop}%
\bibitem [{\citenamefont {Sallamie}\ and\ \citenamefont
  {Shaw}(2005)}]{Sallamie2005}%
  \BibitemOpen
  \bibfield  {author} {\bibinfo {author} {\bibfnamefont {N.}~\bibnamefont
  {Sallamie}}\ and\ \bibinfo {author} {\bibfnamefont {J.}~\bibnamefont
  {Shaw}},\ }\href {\doibase 10.1016/j.fluid.2005.07.022} {\bibfield  {journal}
  {\bibinfo  {journal} {Fluid Phase Equilibria}\ }\textbf {\bibinfo {volume}
  {237}},\ \bibinfo {pages} {100} (\bibinfo {year} {2005})}\BibitemShut
  {NoStop}%
\bibitem [{\citenamefont {Wei}\ \emph {et~al.}(1979)\citenamefont {Wei},
  \citenamefont {Kalyanaraman}, \citenamefont {Singer},\ and\ \citenamefont
  {Garito}}]{Wei1979}%
  \BibitemOpen
  \bibfield  {author} {\bibinfo {author} {\bibfnamefont {T.}~\bibnamefont
  {Wei}}, \bibinfo {author} {\bibfnamefont {P.~S.}\ \bibnamefont
  {Kalyanaraman}}, \bibinfo {author} {\bibfnamefont {K.~D.}\ \bibnamefont
  {Singer}}, \ and\ \bibinfo {author} {\bibfnamefont {A.~F.}\ \bibnamefont
  {Garito}},\ }\href {\doibase 10.1103/PhysRevB.20.5090} {\bibfield  {journal}
  {\bibinfo  {journal} {Physical Review B}\ }\textbf {\bibinfo {volume} {20}},\
  \bibinfo {pages} {5090} (\bibinfo {year} {1979})}\BibitemShut {NoStop}%
\bibitem [{\citenamefont {Sekine}\ \emph {et~al.}(1996)\citenamefont {Sekine},
  \citenamefont {Tsutsumi},\ and\ \citenamefont {Tanokura}}]{Sekine1996}%
  \BibitemOpen
  \bibfield  {author} {\bibinfo {author} {\bibfnamefont {T.}~\bibnamefont
  {Sekine}}, \bibinfo {author} {\bibfnamefont {K.}~\bibnamefont {Tsutsumi}}, \
  and\ \bibinfo {author} {\bibfnamefont {Y.}~\bibnamefont {Tanokura}},\ }\href
  {http://www.sciencedirect.com/science/article/pii/0921452695008020}
  {\bibfield  {journal} {\bibinfo  {journal} {Physica B: Condensed Matter}\
  }\textbf {\bibinfo {volume} {220}},\ \bibinfo {pages} {532} (\bibinfo {year}
  {1996})}\BibitemShut {NoStop}%
\bibitem [{\citenamefont {Dlott}(1986)}]{Dlott1986}%
  \BibitemOpen
  \bibfield  {author} {\bibinfo {author} {\bibfnamefont {D.~D.}\ \bibnamefont
  {Dlott}},\ }\href {\doibase 10.1146/annurev.pc.37.100186.001105} {\bibfield
  {journal} {\bibinfo  {journal} {Annual Review of Physical Chemistry}\
  }\textbf {\bibinfo {volume} {37}},\ \bibinfo {pages} {157} (\bibinfo {year}
  {1986})}\BibitemShut {NoStop}%
\bibitem [{\citenamefont {Galchenkov}\ \emph {et~al.}(1998)\citenamefont
  {Galchenkov}, \citenamefont {Ivanov}, \citenamefont {Pyataikin},
  \citenamefont {Chernov},\ and\ \citenamefont {Monceau}}]{Galchenkov1998}%
  \BibitemOpen
  \bibfield  {author} {\bibinfo {author} {\bibfnamefont {L.~A.}\ \bibnamefont
  {Galchenkov}}, \bibinfo {author} {\bibfnamefont {S.~N.}\ \bibnamefont
  {Ivanov}}, \bibinfo {author} {\bibfnamefont {I.~I.}\ \bibnamefont
  {Pyataikin}}, \bibinfo {author} {\bibfnamefont {V.~P.}\ \bibnamefont
  {Chernov}}, \ and\ \bibinfo {author} {\bibfnamefont {P.}~\bibnamefont
  {Monceau}},\ }\href {\doibase 10.1103/PhysRevB.57.13220} {\bibfield
  {journal} {\bibinfo  {journal} {Physical Review B}\ }\textbf {\bibinfo
  {volume} {57}},\ \bibinfo {pages} {13220} (\bibinfo {year}
  {1998})}\BibitemShut {NoStop}%
\bibitem [{\citenamefont {M\"{u}ller}\ \emph {et~al.}(2009)\citenamefont
  {M\"{u}ller}, \citenamefont {Brandenburg},\ and\ \citenamefont
  {Schlueter}}]{Mueller2009}%
  \BibitemOpen
  \bibfield  {author} {\bibinfo {author} {\bibfnamefont {J.}~\bibnamefont
  {M\"{u}ller}}, \bibinfo {author} {\bibfnamefont {J.}~\bibnamefont
  {Brandenburg}}, \ and\ \bibinfo {author} {\bibfnamefont {J.~A.}\ \bibnamefont
  {Schlueter}},\ }\href {\doibase 10.1103/PhysRevB.79.214521} {\bibfield
  {journal} {\bibinfo  {journal} {Phys. Rev. B}\ }\textbf {\bibinfo {volume}
  {79}},\ \bibinfo {pages} {214521} (\bibinfo {year} {2009})}\BibitemShut
  {NoStop}%
\bibitem [{\citenamefont {Pop}\ \emph {et~al.}(2005)\citenamefont {Pop},
  \citenamefont {Mann}, \citenamefont {Cao}, \citenamefont {Wang},
  \citenamefont {Goodson},\ and\ \citenamefont {Dai}}]{Pop2005}%
  \BibitemOpen
  \bibfield  {author} {\bibinfo {author} {\bibfnamefont {E.}~\bibnamefont
  {Pop}}, \bibinfo {author} {\bibfnamefont {D.}~\bibnamefont {Mann}}, \bibinfo
  {author} {\bibfnamefont {J.}~\bibnamefont {Cao}}, \bibinfo {author}
  {\bibfnamefont {Q.}~\bibnamefont {Wang}}, \bibinfo {author} {\bibfnamefont
  {K.}~\bibnamefont {Goodson}}, \ and\ \bibinfo {author} {\bibfnamefont
  {H.}~\bibnamefont {Dai}},\ }\href {\doibase 10.1103/PhysRevLett.95.155505}
  {\bibfield  {journal} {\bibinfo  {journal} {Physical Review Letters}\
  }\textbf {\bibinfo {volume} {95}},\ \bibinfo {pages} {155505} (\bibinfo
  {year} {2005})}\BibitemShut {NoStop}%
\bibitem [{\citenamefont {Berciaud}\ \emph {et~al.}(2010)\citenamefont
  {Berciaud}, \citenamefont {Han}, \citenamefont {Mak}, \citenamefont {Brus},
  \citenamefont {Kim},\ and\ \citenamefont {Heinz}}]{Berciaud2010}%
  \BibitemOpen
  \bibfield  {author} {\bibinfo {author} {\bibfnamefont {S.}~\bibnamefont
  {Berciaud}}, \bibinfo {author} {\bibfnamefont {M.~Y.}\ \bibnamefont {Han}},
  \bibinfo {author} {\bibfnamefont {K.~F.}\ \bibnamefont {Mak}}, \bibinfo
  {author} {\bibfnamefont {L.~E.}\ \bibnamefont {Brus}}, \bibinfo {author}
  {\bibfnamefont {P.}~\bibnamefont {Kim}}, \ and\ \bibinfo {author}
  {\bibfnamefont {T.~F.}\ \bibnamefont {Heinz}},\ }\href {\doibase
  10.1103/PhysRevLett.104.227401} {\bibfield  {journal} {\bibinfo  {journal}
  {Physical Review Letters}\ }\textbf {\bibinfo {volume} {104}},\ \bibinfo
  {pages} {227401} (\bibinfo {year} {2010})}\BibitemShut {NoStop}%
\bibitem [{\citenamefont {Hill}\ \emph {et~al.}(1996)\citenamefont {Hill},
  \citenamefont {Sandhu}, \citenamefont {Boonman}, \citenamefont {Perenboom},
  \citenamefont {Wittlin}, \citenamefont {Uji}, \citenamefont {Brooks},
  \citenamefont {Kato}, \citenamefont {Sawa},\ and\ \citenamefont
  {Aonuma}}]{Hill1996}%
  \BibitemOpen
  \bibfield  {author} {\bibinfo {author} {\bibfnamefont {S.}~\bibnamefont
  {Hill}}, \bibinfo {author} {\bibfnamefont {P.~S.}\ \bibnamefont {Sandhu}},
  \bibinfo {author} {\bibfnamefont {M.~E.~J.}\ \bibnamefont {Boonman}},
  \bibinfo {author} {\bibfnamefont {J.~A. A.~J.}\ \bibnamefont {Perenboom}},
  \bibinfo {author} {\bibfnamefont {A.}~\bibnamefont {Wittlin}}, \bibinfo
  {author} {\bibfnamefont {S.}~\bibnamefont {Uji}}, \bibinfo {author}
  {\bibfnamefont {J.~S.}\ \bibnamefont {Brooks}}, \bibinfo {author}
  {\bibfnamefont {R.}~\bibnamefont {Kato}}, \bibinfo {author} {\bibfnamefont
  {H.}~\bibnamefont {Sawa}}, \ and\ \bibinfo {author} {\bibfnamefont
  {S.}~\bibnamefont {Aonuma}},\ }\href {\doibase 10.1103/PhysRevB.54.13536}
  {\bibfield  {journal} {\bibinfo  {journal} {Physical Review B}\ }\textbf
  {\bibinfo {volume} {54}},\ \bibinfo {pages} {13536} (\bibinfo {year}
  {1996})}\BibitemShut {NoStop}%
\bibitem [{\citenamefont {Chaikin}\ \emph {et~al.}(1981)\citenamefont
  {Chaikin}, \citenamefont {Haen}, \citenamefont {Engler},\ and\ \citenamefont
  {Greene}}]{Chaikin1981}%
  \BibitemOpen
  \bibfield  {author} {\bibinfo {author} {\bibfnamefont {P.~M.}\ \bibnamefont
  {Chaikin}}, \bibinfo {author} {\bibfnamefont {P.}~\bibnamefont {Haen}},
  \bibinfo {author} {\bibfnamefont {E.~M.}\ \bibnamefont {Engler}}, \ and\
  \bibinfo {author} {\bibfnamefont {R.~L.}\ \bibnamefont {Greene}},\ }\href
  {\doibase 10.1103/PhysRevB.24.7155} {\bibfield  {journal} {\bibinfo
  {journal} {Physical Review B}\ }\textbf {\bibinfo {volume} {24}},\ \bibinfo
  {pages} {7155} (\bibinfo {year} {1981})}\BibitemShut {NoStop}%
\bibitem [{\citenamefont {Su}\ and\ \citenamefont {Schrieffer}(1980)}]{Su1980}%
  \BibitemOpen
  \bibfield  {author} {\bibinfo {author} {\bibfnamefont {W.~P.}\ \bibnamefont
  {Su}}\ and\ \bibinfo {author} {\bibfnamefont {J.~R.}\ \bibnamefont
  {Schrieffer}},\ }\href {\doibase 10.1073/pnas.77.10.5626} {\bibfield
  {journal} {\bibinfo  {journal} {Proceedings of the National Academy of
  Sciences}\ }\textbf {\bibinfo {volume} {77}},\ \bibinfo {pages} {5626}
  (\bibinfo {year} {1980})}\BibitemShut {NoStop}%
\bibitem [{\citenamefont {Ivek}\ \emph {et~al.}(2010)\citenamefont {Ivek},
  \citenamefont {Korin-Hamzi\'{c}}, \citenamefont {Milat}, \citenamefont
  {Tomi\'{c}}, \citenamefont {Clauss}, \citenamefont {Drichko}, \citenamefont
  {Schweitzer},\ and\ \citenamefont {Dressel}}]{Ivek2010}%
  \BibitemOpen
  \bibfield  {author} {\bibinfo {author} {\bibfnamefont {T.}~\bibnamefont
  {Ivek}}, \bibinfo {author} {\bibfnamefont {B.}~\bibnamefont
  {Korin-Hamzi\'{c}}}, \bibinfo {author} {\bibfnamefont {O.}~\bibnamefont
  {Milat}}, \bibinfo {author} {\bibfnamefont {S.}~\bibnamefont {Tomi\'{c}}},
  \bibinfo {author} {\bibfnamefont {C.}~\bibnamefont {Clauss}}, \bibinfo
  {author} {\bibfnamefont {N.}~\bibnamefont {Drichko}}, \bibinfo {author}
  {\bibfnamefont {D.}~\bibnamefont {Schweitzer}}, \ and\ \bibinfo {author}
  {\bibfnamefont {M.}~\bibnamefont {Dressel}},\ }\href {\doibase
  10.1103/PhysRevLett.104.206406} {\bibfield  {journal} {\bibinfo  {journal}
  {Physical Review Letters}\ }\textbf {\bibinfo {volume} {104}},\ \bibinfo
  {pages} {206406} (\bibinfo {year} {2010})}\BibitemShut {NoStop}%
\bibitem [{\citenamefont {Maki}(1982)}]{Maki1982}%
  \BibitemOpen
  \bibfield  {author} {\bibinfo {author} {\bibfnamefont {K.}~\bibnamefont
  {Maki}},\ }\href {\doibase 10.1103/PhysRevB.26.2181} {\bibfield  {journal}
  {\bibinfo  {journal} {Physical Review B}\ }\textbf {\bibinfo {volume} {26}},\
  \bibinfo {pages} {2181} (\bibinfo {year} {1982})}\BibitemShut {NoStop}%
\bibitem [{\citenamefont {Kreith}\ \emph {et~al.}(2010)\citenamefont {Kreith},
  \citenamefont {Manglik},\ and\ \citenamefont {Bohn}}]{Kreith2010}%
  \BibitemOpen
  \bibfield  {author} {\bibinfo {author} {\bibfnamefont {F.}~\bibnamefont
  {Kreith}}, \bibinfo {author} {\bibfnamefont {R.}~\bibnamefont {Manglik}}, \
  and\ \bibinfo {author} {\bibfnamefont {M.}~\bibnamefont {Bohn}},\ }\href
  {https://books.google.de/books?hl=de\&lr=\&id=2TYKAAAAQBAJ\&oi=fnd\&pg=PR7\&dq=Principles+of+Heat+Transfer+kreith\&ots=zGLwata87A\&sig=BJBugfz5lK0y17Db9D3rdjQT2Wg}
  {\emph {\bibinfo {title} {{Principles of Heat Transfer}}}}\ (\bibinfo
  {publisher} {Cengage Learning},\ \bibinfo {year} {2010})\BibitemShut
  {NoStop}%
\bibitem [{\citenamefont {Brill}\ \emph {et~al.}(1981)\citenamefont {Brill},
  \citenamefont {Tzou}, \citenamefont {Verma},\ and\ \citenamefont
  {Ong}}]{Brill1981}%
  \BibitemOpen
  \bibfield  {author} {\bibinfo {author} {\bibfnamefont {J.}~\bibnamefont
  {Brill}}, \bibinfo {author} {\bibfnamefont {C.}~\bibnamefont {Tzou}},
  \bibinfo {author} {\bibfnamefont {G.}~\bibnamefont {Verma}}, \ and\ \bibinfo
  {author} {\bibfnamefont {N.}~\bibnamefont {Ong}},\ }\href {\doibase
  10.1016/0038-1098(81)90663-3} {\bibfield  {journal} {\bibinfo  {journal}
  {Solid State Communications}\ }\textbf {\bibinfo {volume} {39}},\ \bibinfo
  {pages} {233} (\bibinfo {year} {1981})}\BibitemShut {NoStop}%
\bibitem [{\citenamefont {Torizuka}\ \emph {et~al.}(2005)\citenamefont
  {Torizuka}, \citenamefont {Tajima},\ and\ \citenamefont
  {Yamamoto}}]{Torizuka2005a}%
  \BibitemOpen
  \bibfield  {author} {\bibinfo {author} {\bibfnamefont {K.}~\bibnamefont
  {Torizuka}}, \bibinfo {author} {\bibfnamefont {H.}~\bibnamefont {Tajima}}, \
  and\ \bibinfo {author} {\bibfnamefont {T.}~\bibnamefont {Yamamoto}},\ }\href
  {\doibase 10.1103/PhysRevB.71.193101} {\bibfield  {journal} {\bibinfo
  {journal} {Physical Review B}\ }\textbf {\bibinfo {volume} {71}},\ \bibinfo
  {pages} {193101} (\bibinfo {year} {2005})}\BibitemShut {NoStop}%
\end{thebibliography}%

\end{document}